\documentclass[twocolumn,twoside]{revtex4}
\usepackage{graphicx}
\usepackage{fancyhdr}
\begin{document}

\title{
Time-dependent CP violation in $B_s^0$ decays at LHCb}


\author{E. P. M. Gabriel\\
on behalf of the LHCb Collaboration}
\affiliation{University of Edinburgh, Edinburgh EH8 9YL, United Kingdom}

\begin{abstract}
The $B_s^0-\bar{B}_s^0$ system can be used to look for new sources of CP violation. Time-dependent CP violation measurements of beauty mesons allow the determination of the mixing induced CP-violating phase, $\phi_s$. The angular analysis of the $B_s^0\rightarrow \phi \phi$ decay is presented, along with two $B_s^0\rightarrow J/\psi h^+h^-$ decays, where the two hadrons are either a pair of kaons or a pair of pions. The analyses are based on data collected with the LHCb detector between $2011$ and $2016$.

\end{abstract}

\maketitle

\thispagestyle{fancy}


%%%%%%%%%%%%%%%%%%%%%%%%%%%%%%%%%%%%%%%%%%%%%%%%%%%%%%%%%%%%%
\section{Introduction}

The Standard Model (SM) fails to explain the matter-antimatter asymmetry observed in our universe. Finding new sources of CP violation could aid
in explaining this difference. Time-dependent CP violation
measurements of beauty mesons allow the determination of the
mixing induced CP-violating phase, $\phi_s$. The CP-violating
phase is of interest in both penguin dominated and three-level $b\rightarrow s$
transitions, which
test the flavour changing neutral current interaction
describing B meson mixing. The LHCb experiment provides high
sensitivity in these measurements.

We present new results of time-dependent CP violation in the
$B_s^0-\bar{B}_s^0$ system using data collected at LHCb between
2011 and 2016. The analyses discussed are the time-dependent
analysis of the $B_s^0\rightarrow \phi \phi$ decay, as well as
two separate $B_s^0\rightarrow J/\psi h^+h^-$ decays, where the two
hadrons are either a pair of kaons or a pair of pions.

%%%%%%%%%%%%%%%%%%%%%%%%%%%%%%%%%%%%%%%%%%%%%%%%%%%%%%%%%%%%%
\section{Motivation}
\label{sec:motivation}

In $B_s^0$ decays that proceed through a $b\rightarrow c\bar{c}s$ transition,
the CP-violating phase, $\phi_s^{c\bar{c}s}$ is given by
$-2\beta_s$ where higher order diagrams including loops and new
physics (NP) contributions are neglected. This is analogous to
the CKM-angle $\beta$ in $B^0$ decays. The CKM angle
$\beta_s$ is given by 
\begin{equation}
    \beta_s\equiv \mathrm{arg} \left(\frac{-V_{ts}V_{tb}^*}{V_{cs}V_{cb}^*}\right),
\end{equation}
where the arguments are elements from the 
Cabibbo Kobayashi Maskawa (CKM) matrix. 

\begin{figure}[h]
\centering
\includegraphics[width=80mm]{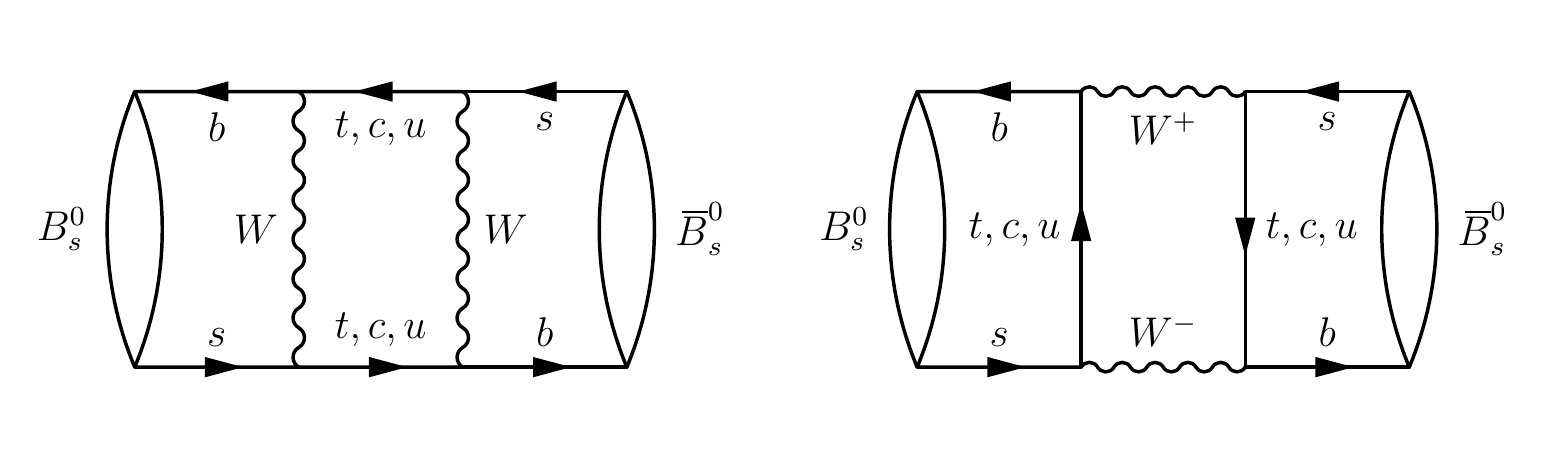}
\caption{The box diagrams representing $B_s^0-\bar{B}_s^0$ mixing.} 
\label{fig:box}
\end{figure}

\begin{figure}[h]
\centering
\includegraphics[width=40mm]{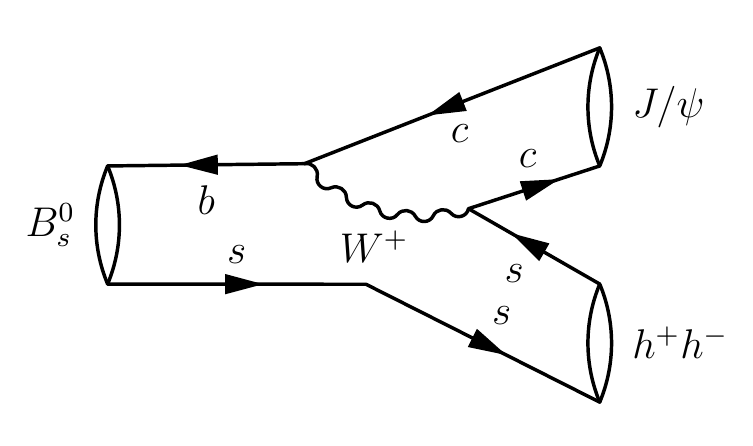}
\caption{The tree diagram of the $B_s^0\rightarrow J/\psi h^+h^-$ decay.} 
\label{fig:tree}
\end{figure}

The numerator arguments, $V_{ts}$ and $V_{tb}$, stem from the
$B_s^0-\bar{B}_s^0$ mixing Feynman diagram shown in Fig.~\ref{fig:box}. The two matrix
elements in the denominator enter through the tree diagram of 
a $b\rightarrow c\bar{c}s$ transition, as can be seen in Fig.~\ref{fig:tree}.
Thus, the CP-violating phase, $\phi_s^{c\bar{c}s}$, can be measured through interference between mixing and decay as $\phi_s \equiv \phi_{\mathrm{mix}}-2\phi_{\mathrm{dec}}$.

%%%%%%%%%%%%%%%%%%%%%%%%%%%%%%%%%%%%%%%%%%%%%%%%%%%%%%%%%%%%%
\section{Status and predictions of the decays}

The updated measurement of CP violation parameters in the $B_s^0\rightarrow
\phi \phi$ decay has been performed on the full Run 1 data-set
with the addition of data collected in 2015 and
2016~\cite{LHCb-PAPER-2019-019}. The decay mode is dominated by penguin loop
contributions, thus increasing the sensitivity to NP. In the
analysis, the direct CP violation parameter, $|\lambda|$ is
measured, along with the CP-violating phase
$\phi_s^{s\bar{s}s}$. 
%Note that this phase is different from the phase measured in the $B_s^0\rightarrow J/\psi h^+h^-$ decay
The SM predicts $\phi_s^{s\bar{s}s}$ close to zero in the
context of QCD factorisation~\cite{PhysRevD.80.114026}. Theoretical errors are of the
order of $\sim 2 \%$\cite{Bartsch:2008ps}. However, several beyond
the Standard Model (BSM) models allow for significant CP
violation in $b\rightarrow s\bar{s}s$ penguin decays~\cite{MOROI2000366, Nandi2006, DATTA2009256}.

The $B_s^0\rightarrow J/\psi h^+h^-$ analyses are both
performed using data collected by the LHCb detector in 2015 
and 2016. The results have been combined with previous Run 1
measurements~\cite{LHCb-PAPER-2014-059, LHCb-PAPER-2014-019}. The common parameters measured here are
$|\lambda|$ and the CP-violating phase $\phi_s^{c\bar{c}s}$.

The $B_s^0\rightarrow J/\psi K^+K^-$ analysis~\cite{LHCb-PAPER-2019-013} focuses on the
$\phi$ mass window by selecting the $K^+K^-$ invariant mass 
between $0.99$ GeV/$c^2$ and $1.05$ GeV/$c^2$. Additional
parameters measured in this decay are the decay width
difference, $\Delta \Gamma_s$, as well as the difference between average decay widths, $\Gamma_s -
\Gamma_d$.

\begin{figure}[h]
\centering
\includegraphics[width=80mm]{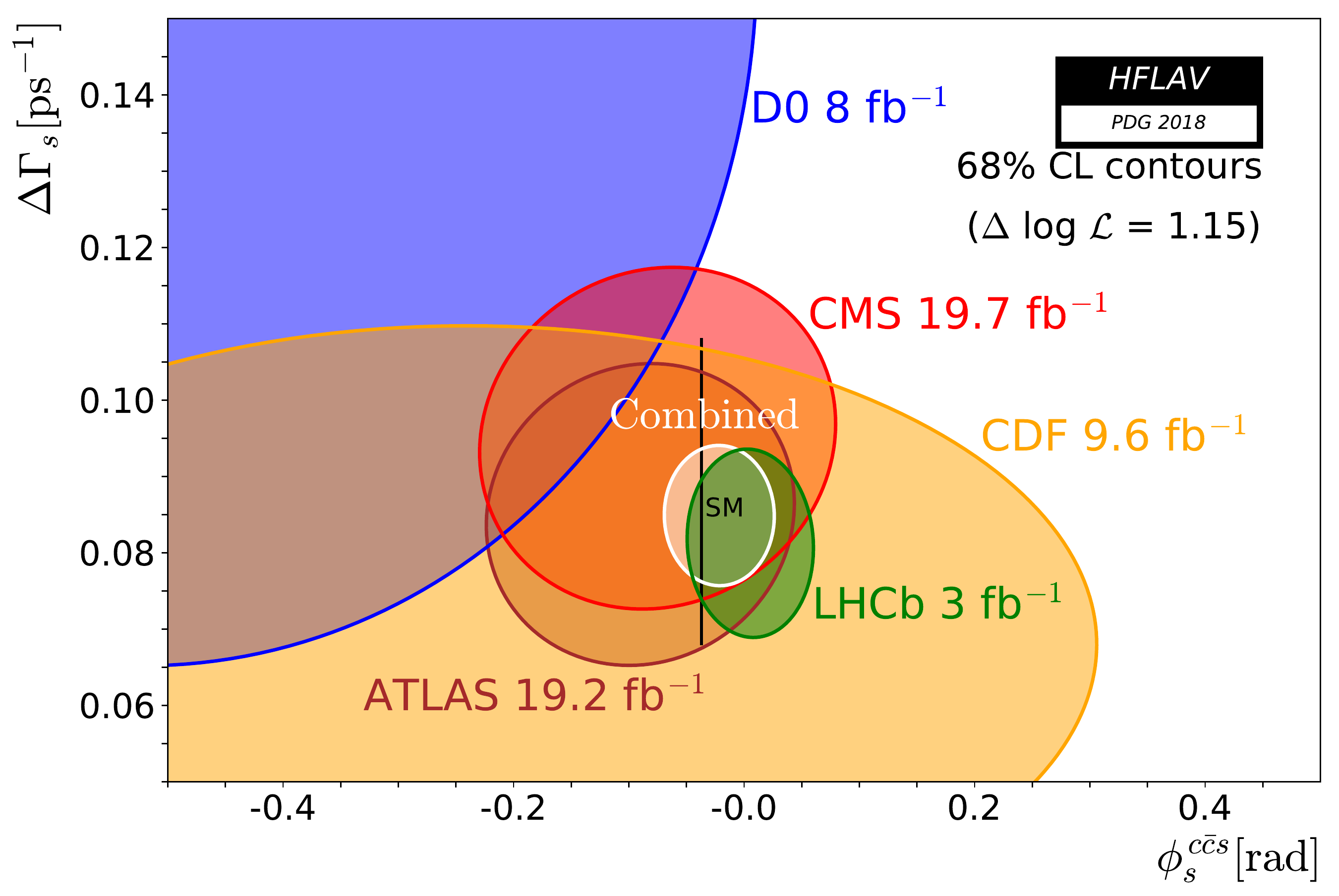}
\caption{Experimental status of $\phi_s^{c\bar{c}s}$ vs. $\Delta \Gamma_s$ before the Winter 2019 conferences~\cite{HFLAV2018}.} 
\label{fig:prewinter}
\end{figure}

The $B_s^0\rightarrow J/\psi \pi^+\pi^-$ decay has previously
been shown to have a predominant CP-odd final state~\cite{LHCb-PAPER-2013-069}.
This implies the decay has no sensitivity to $\Gamma_L$, such that the measured difference in average decay widths is $\Gamma_H -
\Gamma_d$. 
Predictions from global fits to data of the phase $\phi_s^{c\bar{c}s}$ give a value of $-36.8
^{+0.7}_{-1.0}$ [mrad]~\cite{CKMfitter}. Its experimental status
before the Winter 2019 conferences is shown in
Fig~\ref{fig:prewinter}.

%%%%%%%%%%%%%%%%%%%%%%%%%%%%%%%%%%%%%%%%%%%%%%%%%%%%%%%%%%%%%
\section{Analysis ingredients}

The analyses discussed here are performed using a
time-dependent angular analysis. The methods
used are similar and a general overview of the necessary analysis ingredients is described in this section.

\subsection{Selection}

It is essential to select a sample of events with signal purity
as high as possible. Each analysis uses different methods to
achieve this. In the $B_s^0\rightarrow \phi \phi$ analysis, a
multivariate neural network is trained to remove background
events. The $\Lambda_b^0\rightarrow \phi p K^-$ background is
modeled in the fit to the data due to it being difficult to
remove. Roughly 8500 signal candidates were found in the full
2011-2016 data-set. 

The $B_s^0\rightarrow J/\psi h^+h^-$ analyses use a
boosted decision tree to select signal events. In the
$B_s^0\rightarrow J/\psi K^+K^-$ decay mode, the $
\Lambda_b^0\rightarrow J\psi p K^-$ background mode is
subtracted using negatively weighted MC candidates. A total of
around 117 000 signal events are found.
%as seen in Fig~\ref{fig:jpsiphimass}. 
The $B_s^0\rightarrow
J/\psi \pi^+\pi^-$ analysis uses the wrong sign
($B_s^0\rightarrow J/\psi \pi^{\pm}\pi^{\pm}$) invariant mass
shape to determine the shape of the combinatorial background.
This shape is used in the final mass fit, which yields roughly
33 500 signal candidates.

%\begin{figure}[h]
%\centering
%\includegraphics[width=80mm]{jpsiphimass.png}
%\caption{A fit to the $J/\psi K^+K^-$ invariant mass distribution selecting $B_s^0\rightarrow J/\psi K^+K^-$ candidates.} 
%\label{fig:jpsiphimass}
%\end{figure}

\subsection{Decay-time resolution}

To resolve the fast flavour oscillations induced by $B_s^0-\bar{B}_s^0$ meson mixing, it is essential to achieve a good decay-time resolution that is much smaller than the oscillation period. In the analyses covered here, an average decay-time resolution of $41-45$ fs is accomplished. 

The $B_s^0\rightarrow J/\psi h^+h^-$ analyses make use of a prompt $J/\psi$ sample to calibrate the decay-time resolution. In the $B_s^0\rightarrow \phi \phi$ analysis, a novel method was implemented, utilising the fact that the opening angle between the two kaons stemming from a $\phi$ is very small. This is exploited by calibrating the decay-time resolution on a prompt pseudo-two body sample.  

\subsection{Angular efficiency}

\begin{figure}[h]
\centering
\includegraphics[width=80mm]{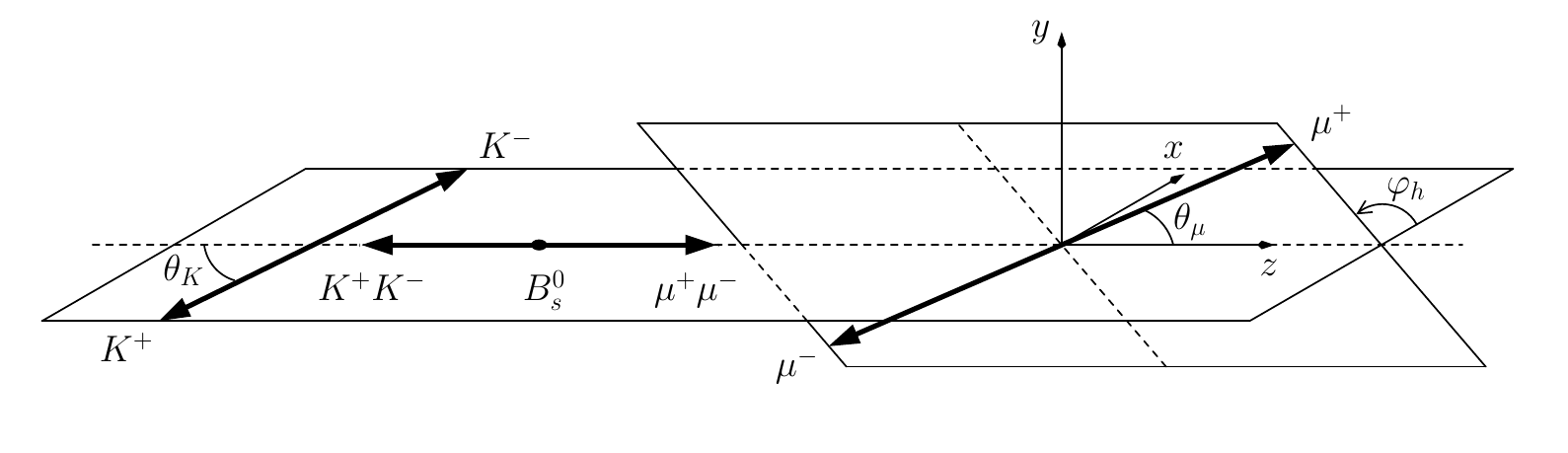}
\caption{Definition of the helicity angles used in the time-dependent angular $B_s^0\rightarrow J/\psi K^+K^-$ analyses. } 
\label{fig:helicity}
\end{figure}

The non-uniform selection efficiency in the decay angles as a result of the LHCb detector acceptance and kinematic selection cuts needs to be accounted for. The helicity angles used in the $B_s^0\rightarrow J/\psi K^+K^-$ decay are defined as shown in Fig.~\ref{fig:helicity}. A similar formalism is used in the $B_s^0\rightarrow \phi \phi$ and $B_s^0\rightarrow J/\psi \pi^+\pi^-$ analyses.

The efficiency is calculated for each analysis, where simulated events selected in the same way as data are used. 

\subsection{Decay-time efficiency}

In a similar fashion, the decay-time efficiency needs to be accounted for. All Run 2 (2015-2016) data uses the $B^0\rightarrow J/\psi K^{*0}$ decays as a control mode.
A different control sample is used in Run 1 vs. Run 2 data due to differences in the higher level trigger configuration. In order to create a control sample that is kinematically more similar to the decay mode of interest, the Run 1 $B_s^0\rightarrow \phi \phi$ data use the control mode $B_s^0\rightarrow D_s^-\pi^+$.

\subsection{Flavour tagging}

Knowledge of the flavour of the $B_s^0$ meson at production is essential in measuring the CP-violating phase $\phi_s$, since the sensitivity to the phase scales directly with the effective tagging power. At LHCb, flavour tagging algorithms are generated using self-tagged decay modes, such as $B^+\rightarrow J/\psi K^+$ and $B_s^0\rightarrow D_s^-\pi^+$. The effective tagging power achieved in the analyses discussed here is roughly $5\%$.

%%%%%%%%%%%%%%%%%%%%%%%%%%%%%%%%%%%%%%%%%%%%%%%%%%%%%%%%%%%%%
\section{Results}
\label{sec:results}

The time-dependent fit to extract the CP violation parameters consists of a simultaneous fit to the decay-time and the three helicity angles. The $B_s^0\rightarrow \phi \phi$ fit consists of three contributions: the CP-even $P$-wave and CP-odd $P$-wave coming from the $\phi\phi$ final state. In addition, the $f_0(980)$ resonance is close in mass to the $\phi$ and could thus contribute to the result. This is accounted for in the fit by allowing the combination of an $S$-wave and double $S$-wave component. The fit results are given in Tab.~\ref{tab:phiphiresults}. The results presented are in agreement with previous LHCb results as well as the SM predictions.

\begin{table}[h]
\begin{center}
\caption{$B_s^0\rightarrow \phi \phi$ results}
\begin{tabular}{|c|c|}
\hline \textbf{Parameter} & \textbf{Fit Result}
\\
\hline $\phi_s^{s\bar{s}s}$ & -0.073 $\pm$ 0.115 $\pm$ 0.027 [rad] \\
 $|\lambda|$& 0.99 $\pm$ 0.05 $\pm$ 0.01   \\
\hline
\end{tabular}
\label{tab:phiphiresults}
\end{center}
\end{table}

To extract the fit parameters in the $B_s^0\rightarrow J/\psi K^+K^-$ analysis a similar procedure is used. The difference is that the $m(K^+K^-)$ invariant mass is split into six bins. This aids in controlling the interference of the $\phi$ component with the $S$-wave $f_0(980)$ contribution. The time-dependent angular fit consists of a simultaneous fit in decay-time and helicity angles in the six two-kaon mass bins. The results are shown in Tab.~\ref{tab:jpsiphiresults}. These are the most precise single measurements of $\phi_s^{c\bar{c}s}$, $\Gamma_s - \Gamma_d$ and $\Delta \Gamma_s$. The results are in agreement with SM predictions.

\begin{table}[h]
\begin{center}
\caption{$B_s^0\rightarrow J/\psi K^+K^-$ results}
\begin{tabular}{|c|c|}
\hline \textbf{Parameter} & \textbf{Fit Result}
\\
\hline $\phi_s^{c\bar{c}s}$ & -0.083 $\pm$ 0.041 $\pm$ 0.006 [rad] \\
 $|\lambda|$& 1.012 $\pm$ 0.016 $\pm$ 0.006   \\
 $\Gamma_s - \Gamma_d$ & -0.0041 $\pm$ 0.0024 $\pm$ 0.0015 [ps$^{-1}$] \\
 $\Delta \Gamma_s$ & -0.0772 $\pm$ 0.0077 $\pm$ 0.0026  [ps$^{-1}$] \\
\hline
\end{tabular}
\label{tab:jpsiphiresults}
\end{center}
\end{table}

\begin{figure}[h]
\centering
\includegraphics[width=80mm]{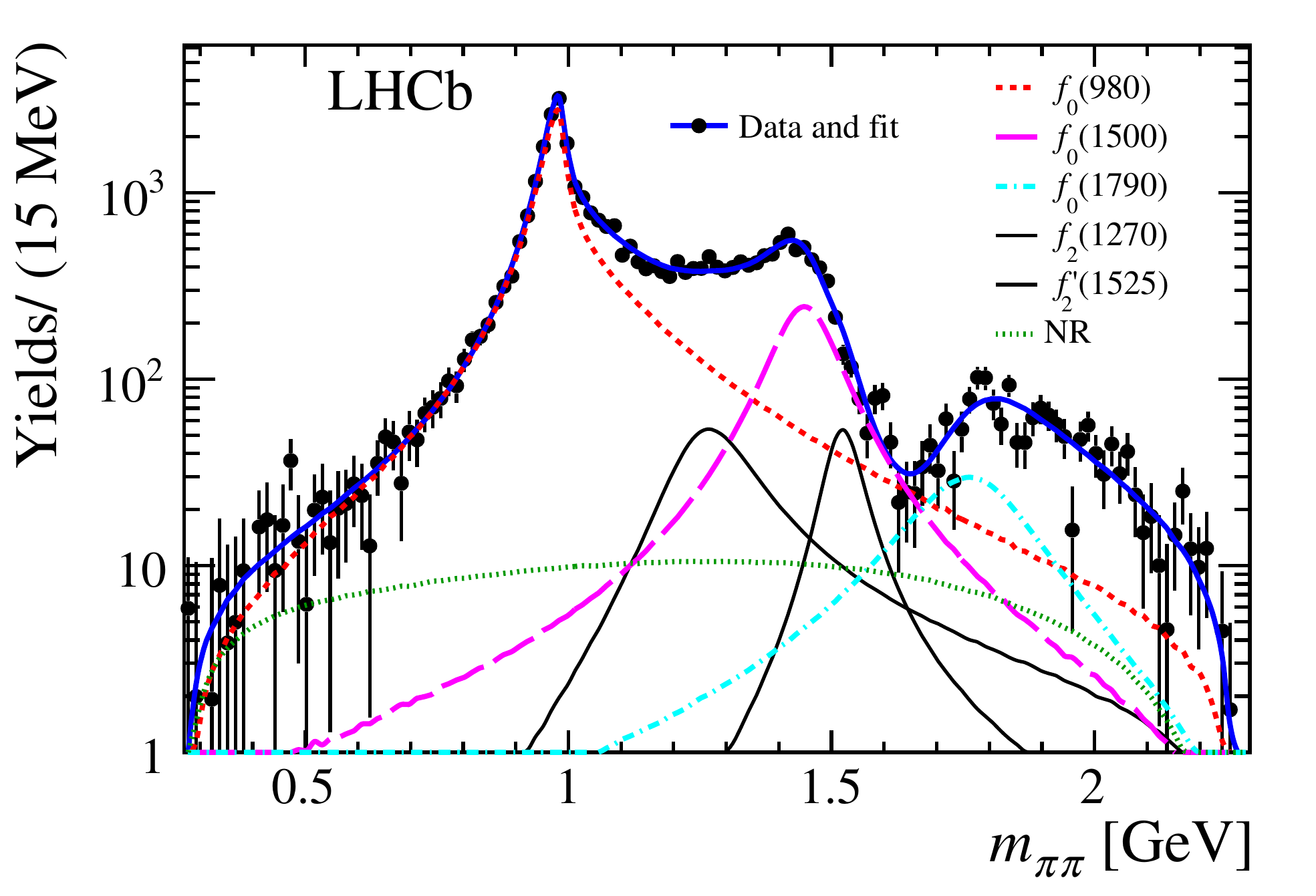}
\caption{A fit to the $\pi^+\pi^-$ invariant mass in the $B_s^0\rightarrow J/\psi \pi^+ \pi^-$ analysis.} 
\label{fig:pipi}
\end{figure}

The $B_s^0\rightarrow J/\psi \pi^+ \pi^-$ analysis performs a
simultaneous fit to the decay-time, helicity angles and the
$\pi^+\pi^-$ mass spectrum. The fit to the $\pi^+\pi^-$
invariant mass, displayed in Fig.~\ref{fig:pipi}, consists of
many resonances. The biggest contribution comes from the
$f_0(980)\rightarrow \pi^+ \pi^-$. Other components in the fit
are the $f_0(1790)$, $f_2(1270)$, $f_2^{'}(1525)$ and
non-resonant contributions. The results are shown in
Tab.~\ref{tab:jpsipipiresults}.

\begin{table}[h]
\begin{center}
\caption{$B_s^0\rightarrow J/\psi \pi^+ \pi^-$ results}
\begin{tabular}{|c|c|}
\hline \textbf{Parameter} & \textbf{Fit Result}
\\
\hline $\phi_s^{c\bar{c}s}$ & -0.057 $\pm$ 0.060 $\pm$ 0.011 [rad] \\
 $|\lambda|$& 1.01$^{+0.08}_{-0.06}$ $\pm$ 0.03   \\
 $\Gamma_H - \Gamma_d$ & -0.050 $\pm$ 0.004 $\pm$ 0.004 [ps$^{-1}$] \\
\hline
\end{tabular}
\label{tab:jpsipipiresults}
\end{center}
\end{table}

%%%%%%%%%%%%%%%%%%%%%%%%%%%%%%%%%%%%%%%%%%%%%%%%%%%%%%%%%%%%%
\section{$\phi_s^{c\bar{c}s}$ combination}
\label{sec:combination}

\begin{figure}[h]
\centering
\includegraphics[width=80mm]{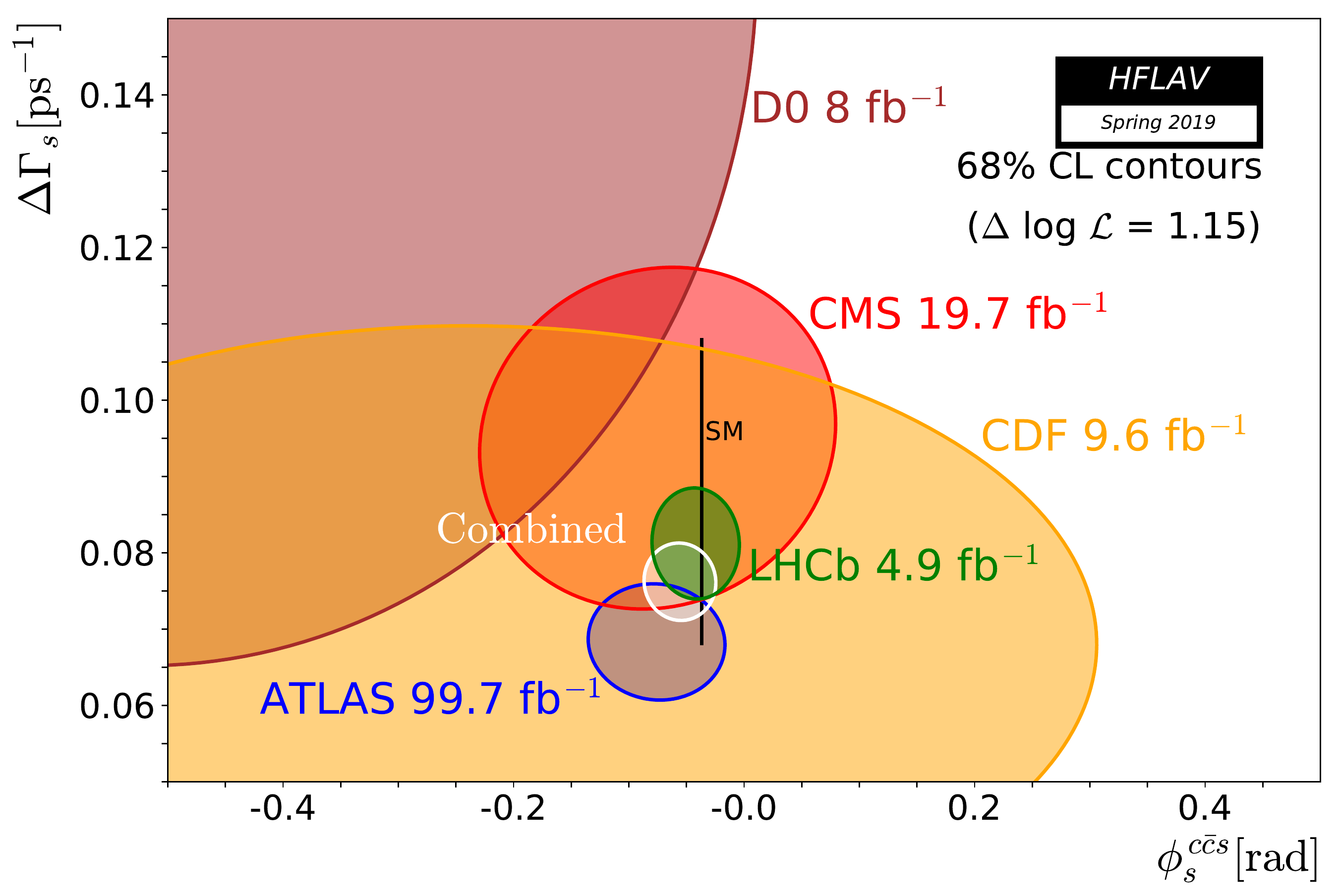}
\caption{Updated experimental status of $\phi_s^{c\bar{c}s}$ vs. $\Delta \Gamma_s$ with the inclusion of the Run 2 $B_s^0\rightarrow J/\psi K^+K^-$ and $B_s^0\rightarrow J/\psi \pi^+\pi^-$ results.} 
\label{fig:postwinter}
\end{figure}

The LHCb collaboration has performed many measurements of the CP-violating phase, $\phi_s^{c\bar{c}s}$, in $b\rightarrow c\bar{c}s$ transitions. The LHCb combined result shown in Fig.~\ref{fig:prewinter} includes results using Run 1 data taken at the LHCb detector from $B_s^0\rightarrow \psi(2S)\phi$~\cite{LHCb-PAPER-2016-027}, $B_s^0\rightarrow D_s^+D_s^-$~\cite{LHCb-PAPER-2014-051}, $B_s^0\rightarrow J/\psi K^+K^-$ where the high $K^+K^-$ mass range is considered ~\cite{LHCb-PAPER-2017-008}, $B_s^0\rightarrow J/\psi K^+K^-$~\cite{LHCb-PAPER-2014-059} and $B_s^0\rightarrow J/\psi \pi^+\pi^-$~\cite{LHCb-PAPER-2014-019}. An updated combination has been performed, including the two new $B_s^0\rightarrow J/\psi h^+h^-$ analyses presented here. The results are displayed in Fig.~\ref{fig:postwinter}. 

The combined results for the $b\rightarrow c\bar{c}s$ parameters are outlined in Tab.~\ref{tab:combinedresults}.
The experimental precision of the measurements presented has improved tremendously with the inclusion of the new measurements.

\begin{table}[h]
\begin{center}
\caption{Updated $b\rightarrow c\bar{c}s$ combination results}
\begin{tabular}{|c|c|}
\hline \textbf{Parameter} & \textbf{Fit Result}
\\
\hline $\phi_s^{c\bar{c}s}$ & -0.041 $\pm$ 0.025 [rad] \\
 $|\lambda|$& 0.993 $\pm$ 0.010    \\
 $\Gamma_s$ & 0.6562 $\pm$ 0.0021 [ps$^{-1}$] \\
 $\Delta \Gamma_s$ & 0.0816 $\pm$ 0.0048 [ps$^{-1}$] \\
\hline
\end{tabular}
\label{tab:combinedresults}
\end{center}
\end{table}

%%%%%%%%%%%%%%%%%%%%%%%%%%%%%%%%%%%%%%%%%%%%%%%%%%%%%%%%%%%%%
\section{Conclusion}
\label{sec:results}

LHCb is currently producing some of the world's most precise measurements in terms of CP-violation in $B_s^0$ meson decays. The results presented do not yet include the data taken at LHCb in $2017$ and $2018$ and it will be very interesting to further increase the precision by including more data.
\begin{figure}[h]
\centering
\includegraphics[width=80mm]{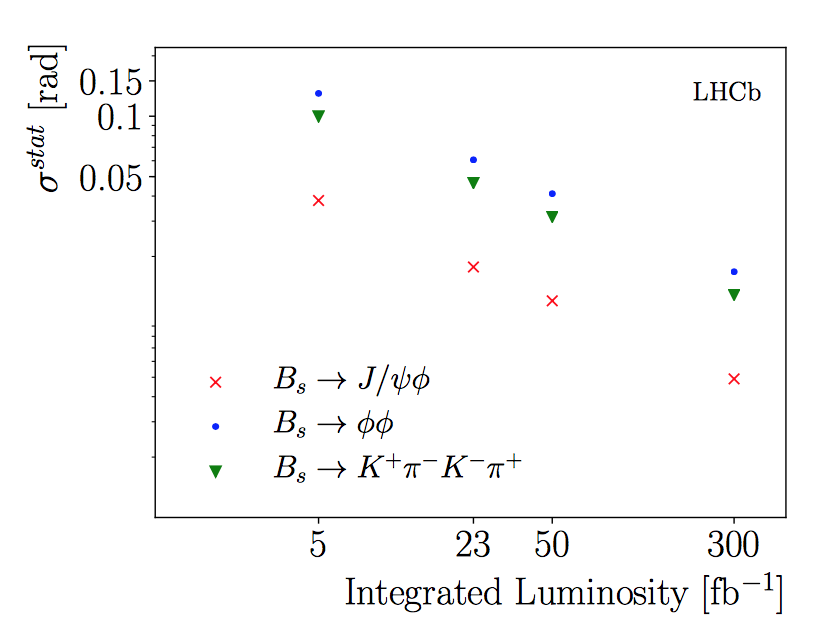}
\caption{Comparison of $\phi_s$ statistical sensitivity from different decay modes.} 
\label{fig:sensitivity}
\end{figure}
The expected increase in sensitivity of $B_s^0\rightarrow \phi \phi$, $B_s^0\rightarrow J/\psi K^+K^-$ and $B_s^0\rightarrow K^+ \pi^- K^- \pi^+$ is shown in Fig.~\ref{fig:sensitivity}, where the current results are based on a sample of data corresponding to $5$fb$^{-1}$.

%\begin{acknowledgments}
%This document is adapted from the ``Instruction for producing FPCP2003
%proceedings'' by P.~Perret 
%and from eConf templates~\cite{templates-ref}.
%\end{acknowledgments}

\bigskip % extra skip inserted
\bibliography{main}

\end{document}